\PassOptionsToPackage{table}{xcolor}
\documentclass[sigconf]{acmart}

\usepackage{xcolor}
\usepackage{enumitem}
\usepackage{subcaption}
\usepackage{booktabs}
\usepackage{multirow}
\usepackage{siunitx}
\usepackage{float}
\usepackage{tcolorbox}
\tcbuselibrary{listings}
\usepackage{graphicx}
\usepackage{array}
\usepackage{colortbl}
\tcbuselibrary{breakable, skins}
\usepackage{tabularx}
\usepackage{soul}        
\sethlcolor{yellow!25}   
\usepackage{pifont}
\usepackage{cuted}
\newcommand{\checkmarkyes}{\ding{51}}

\AtBeginDocument{%
  }

\copyrightyear{2026}
\acmYear{2026}
\setcopyright{cc}
\setcctype{by}
\acmConference[WWW '26]{Proceedings of the ACM Web Conference 2026}{April 13--17, 2026}{Dubai, United Arab Emirates}
\acmBooktitle{Proceedings of the ACM Web Conference 2026 (WWW '26), April 13--17, 2026, Dubai, United Arab Emirates}
\acmDOI{10.1145/3774904.3793060}
\acmISBN{979-8-4007-2307-0/2026/04}

\begin{document}

\title{When Ads Become Profiles: Uncovering the Invisible Risk of Web Advertising at Scale with LLMs}

\author{Baiyu Chen}
\orcid{0009-0000-8617-9635}
\affiliation{%
  \institution{The University of New South Wales}
  \city{Sydney}
  \state{NSW}
  \country{Australia}}
\email{breeze.chen@unsw.edu.au}

\author{Benjamin Tag}
\orcid{0000-0002-7831-2632}
\affiliation{%
  \institution{The University of New South Wales}
  \city{Sydney}
  \state{NSW}
  \country{Australia}}
\email{benjamin.tag@unsw.edu.au}

\author{Hao Xue}
\orcid{0000-0003-1700-9215}
\affiliation{%
  \institution{Hong Kong University of Science and Technology (Guangzhou)}
  \city{Guangzhou}
  \state{Guangdong}
  \country{China}
  }
\email{haoxue@hkust-gz.edu.cn}

\author{Daniel Angus}
\orcid{0000-0002-1412-5096}
\affiliation{%
  \institution{Queensland University of Technology}
  \city{Brisbane}
  \state{QLD}
  \country{Australia}}
\email{daniel.angus@qut.edu.au}

\author{Flora Salim}
\orcid{0000-0002-1237-1664}
\affiliation{%
  \institution{The University of New South Wales}
  \city{Sydney}
  \state{NSW}
  \country{Australia}}
\email{flora.salim@unsw.edu.au}

\renewcommand{\shortauthors}{Baiyu Chen, Benjamin Tag, Hao Xue, Daniel Angus, and Flora Salim}

\begin{abstract}
Regulatory limits on explicit targeting have not eliminated algorithmic profiling on the Web, as optimisation systems still adapt ad delivery to users' private attributes. The widespread availability of powerful zero-shot multimodal Large Language Models (LLMs) has dramatically lowered the barrier for exploiting these latent signals for adversarial inference. We investigate this emerging societal risk, specifically how adversaries can now exploit these signals to reverse-engineer private attributes from ad exposure alone. We introduce a novel pipeline that leverages LLMs as adversarial inference engines to perform natural language profiling. Applying this method to a longitudinal dataset comprising over 435,000 Facebook ad impressions collected from 891 users, we conducted a large-scale study to assess the feasibility and precision of inferring private attributes from passive online ad observations. Our results demonstrate that off-the-shelf LLMs can accurately reconstruct complex user private attributes, including party preference, employment status, and education level, consistently outperforming strong census-based priors and matching or exceeding human social perception at only a fraction of the cost (223× lower) and time (52× faster) required by humans. Critically, actionable profiling is feasible even within short observation windows, indicating that prolonged tracking is not a prerequisite for a successful attack. These findings provide the first empirical evidence that ad streams serve as a high-fidelity digital footprint, enabling off-platform profiling that inherently bypasses current platform safeguards, highlighting a systemic vulnerability in the ad ecosystem and the urgent need for responsible web AI governance in the generative AI era. The code is available at \url{https://github.com/Breezelled/when-ads-become-profiles}.
\end{abstract}

\ccsdesc[500]{Information systems~Online advertising}
\ccsdesc[500]{Security and privacy~Social aspects of security and privacy}
\ccsdesc[300]{Computing methodologies~Artificial intelligence}
\keywords{Targeted Advertising; Large Language Models; Privacy Risks}


\maketitle

\section{Introduction}
\label{sec:intro}

\begin{figure*}[htbp]
    \centering
    \includegraphics[width=0.707\linewidth]{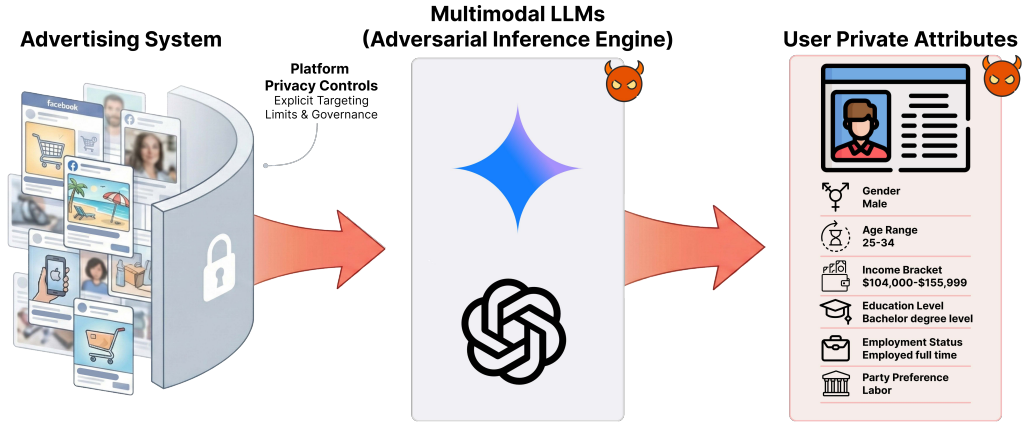}
    \caption{Conceptual overview of the adversarial profiling threat from the passive ad exposure to the user.}
    \label{fig:overview}
\end{figure*}

Online advertising serves as the primary economic engine of the Web, enabling the free distribution of content and services. To maximise relevance and revenue, platforms like Facebook and Google deploy sophisticated, opaque algorithms to deliver personalised content, matching advertisements to users based on granular inferred profiles and behaviours~\cite{Computational_Advertising_Recent_Advances}. 
However, as privacy concerns have mounted, the Web ecosystem has shifted toward stricter governance and regulatory compliance~\cite{GDPR2016}. A pivotal moment occurred in 2022, when major platforms like Meta removed "Detailed Targeting" options for sensitive categories, including political beliefs, health causes, sexual orientation, and religious practices, to prevent potential abuse and discrimination~\cite{meta2022removing}. This shift was intended to enhance user privacy by limiting how advertisers could directly target vulnerable groups through the platform's native tools.

While these policies limit explicit targeting options, they do not neutralise the demographic skews introduced by the platform's optimisation algorithms. Prior studies have established that ad delivery remains highly correlated with private attributes, even under neutral targeting settings~\cite{ali2019discrimination, imana2021job, ali2022all,ali2023problematic}, as an artifact of algorithmic optimisation for user engagement. Our own preliminary analysis confirms that these correlations persist within the dataset we use, creating a discernible link between ad exposure and user private attributes. Consequently, the stream of advertisements shown to a user continues to serve as a high-fidelity proxy for their identity, containing latent fine-grained signals about their private attributes.

The emergence of generative AI challenges the efficacy of existing governance frameworks, introducing potential vulnerabilities by diversifying attack pathways and increasing uncertainty.
Previously, exploiting these signals required significant resources: collecting massive labeled datasets and training classifiers~\cite{bi2013inferring,kosinski2013private,schwartz2013personality,hinds2018demographic, tricomi2024spotify}. However, this barrier has been fundamentally altered by the rapid democratisation of powerful generative AI like Large Language Models (LLMs). Such technologies demonstrate remarkable abilities to understand and generate nuanced, human-like content~\cite{hurst2024gpt, openai2025gpt5systemcard}, reason about complex contexts~\cite{jaech2024openai,guo2025deepseek}, adapt to user-specific preferences~\cite{zhang2025personalization,openai2025memory}, and even integrate multimodal information~\cite{hurst2024gpt,openai2025gpt5systemcard,anthropic2025claude4,comanici2025gemini}, bringing immense convenience but also raising new societal concerns in domains such as online advertising~\cite{AI_Driven_Online_Advertising_Market_Design_Generative_AI_and_Ethics}. As these models become increasingly accessible and capable~\cite{Computational_Advertising_Recent_Advances}, not only through APIs or open-source implementations but also via free, public-facing web interfaces to some of the most advanced proprietary models~\cite{comanici2025gemini, hurst2024gpt, openai2025gpt5systemcard, anthropic2025claude4, guo2025deepseek}, which dramatically lower the barrier for conducting sophisticated analyses at scale, including those with potentially harmful intent. 

Today, even individuals with only basic technical skills can leverage these models to perform such inferences, making the threat of misuse more probable and more accessible. The convergence of latent ad signals and zero-shot capabilities of LLMs creates societal risks: \textbf{regulatory evasion via off-platform profiling enabled by scalable yet imperceptible privacy violations}. Critically, this profiling does not require the distribution of specialised malware, it can be opportunistically deployed within the existing ecosystem of benign-looking browser extensions (e.g., ad blockers, coupon finders), utilising legitimate permissions to silently harvest ad creatives while evading both user suspicion and platform audit trails.

In this work, we present a large-scale Web privacy study to quantify privacy implications of this societal risk. We propose a novel pipeline that leverages the zero-shot capabilities of multimodal LLMs to act as an adversarial inference engine. Utilising a filtered longitudinal dataset collected via a browser extension from 891 Australian Facebook users (spanning over 435,000 individual ad impressions across more than 63,000 sessions), we simulate an attacker's perspective to answer the following: To what extent can multimodal LLMs reconstruct a user's private attributes solely from the sequence of advertisements that are shown to the user?

Our findings reveal that LLMs can reconstruct private attributes such as age, gender, education, employment status and party preference with high accuracy, outperforming strong census-based priors and matching human annotators. Crucially, we demonstrate that this inference is feasible even with short observation windows (session-level), meaning prolonged tracking is not a prerequisite for a successful attack. These findings provide the first empirical demonstration that multimodal LLMs can reverse-engineer user private attributes from passive observation (passive observation here means passively viewed ads from the user's perspective) alone, highlighting the scale of privacy risks and the urgent need for safeguards in the era of accessible, high-capacity generative AI.

\textbf{Our work makes the following contributions:}
\textbf{(1) A Novel Pipeline for Uncovering Emerging Societal Risks in Web Advertising.} We introduce a methodology that leverages multimodal LLMs to perform zero-shot natural language profiling on ad content. Unlike traditional methods that rely on behavioural metadata, our pipeline decodes the semantic signals embedded in visual and textual ad creatives, establishing a new approach for uncovering the privacy dimension of societal risk from unstructured digital footprints. \textbf{(2) Empirical Quantification of Privacy Risk in Passive Ad Streams.} We provide the first large-scale empirical evidence that ad streams serve as a high-fidelity digital footprint. By evaluating LLMs performance against strong census-based priors and human annotators, we demonstrate that LLMs can accurately reconstruct private attributes from passive observation alone, achieving comparable accuracy at a dramatically lower marginal cost (223× cheaper) and latency (52× faster) than human analysis. \textbf{(3) Characterisation of Scalable Profiling.} We reveal the operational viability of this threat by demonstrating that prolonged tracking is unnecessary. We discuss how this capability enables profiling that bypasses governance and regulation, highlighting urgent gaps in current Web privacy protection that are exacerbated as LLMs become increasingly capable and accessible.
\section{Related Works}
\label{sec:related}

In this section, we situate our work at the intersection of digital footprint analysis and the emerging privacy risks posed by LLMs. We review how traditional methods established the link between online behaviour and user attributes, and how recent advancements in LLMs have transformed this landscape from resource-intensive modeling to scalable adversarial inference.

\subsection{Digital Footprints and Traditional Inference}
Prior to the emergence of LLMs, research had established that digital footprints are predictive of private user attributes. Studies have demonstrated that diverse traces, ranging from Facebook Likes~\cite{kosinski2013private} and social media language use~\cite{schwartz2013personality} to search logs~\cite{bi2013inferring} and even public Spotify playlists~\cite{tricomi2024spotify}, contain latent demographic signals. In the visual domain, tasks such as human attribute recognition and pedestrian attribute recognition have achieved high precision in identifying traits like gender and age from images~\cite{app10165608, WANG2022108220}. However, these traditional approaches typically required significant resources, relying on massive labeled datasets to train bespoke classifiers via supervised learning~\cite{hinds2018demographic}.

Within the advertising domain, extensive auditing literature has similarly documented systematic demographic skews in ad delivery algorithms~\cite{ali2019discrimination, imana2021job, ali2022all,ali2023problematic}. While these studies originally aimed to detect algorithmic discrimination, they provide the crucial empirical evidence that ad streams carry latent signals. Our work builds upon this foundation by shifting the lens to adversarial inference, demonstrating how these established correlations can be exploited by off-the-shelf LLMs without the need for the resource-intensive training pipelines required by previous methods.

\subsection{Utility and Risks of LLM Profiling}

The rise of LLMs has fundamentally altered user profiling, shifting from opaque vector representations to transparent, language-based reasoning. In recommender systems, this capability is widely leveraged to construct scrutable user models from interaction histories~\cite{balog2019transparent, radlinski2022natural, ramos-etal-2024-transparent, zhou2024language}, or to infer detailed preferences from sparse inputs like POI check-ins and collaborative signals~\cite{wongso2024genup, ren2024representation, xi2024towards}. Beyond recommendation, LLMs demonstrate strong reasoning capabilities in broader user modeling tasks, ranging from inferring latent interests to generating personalized content~\cite{tan2023user, meguellati2024how}.

Recently, this inferential power has raised significant concerns regarding inferential privacy, where adversaries exploit models to deduce private information from user content. Studies have demonstrated that LLMs can infer personal attributes from online text posts~\cite{staab24beyond}, highlighting that such automated inference operates at a fraction of the cost and time required by human analysts. Extending this to the visual domain, recent works utilise vision-language models to extract attributes from user-uploaded images~\cite{tomekcce2024private} and aggregate clues across personal photo albums using agentic frameworks~\cite{liu2025eye}. Our work identifies a critical gap in this emerging literature. While previous studies focus on \emph{active digital footprints}, content users voluntarily create or upload, we investigate \emph{passive digital footprints}: the stream of advertisements pushed to users by algorithmic systems. By demonstrating that off-the-shelf LLMs can reverse-engineer user private attributes from these passive exposures, we uncover a systemic vulnerability where the platform's own targeting logic can be exploited to compromise user privacy.
\section{Threat Model}
\label{sec:threat_model}

We consider a privacy threat aimed at profiling users based on their passive exposure to algorithmic content. We formalise the adversary's goal, capabilities, and the specific attack vector within the modern Web ecosystem.

\subsection{Adversary Goal and Capabilities}
The adversary aims to infer a target user's private attributes $\mathcal{A}$ (e.g., party preference, employment status) by observing the stream of advertisements $\mathcal{S}$ delivered to them. This adversary seeks to exploit the real-time, high-fidelity targeting intelligence embedded in the ad platform itself. This approach differs fundamentally from existing data acquisition methods. Unlike scraping public social media profiles~\cite{staab24beyond,tomekcce2024private}, which relies on active user disclosure, this attack exploits passive exposure. Furthermore, purchasing data from brokers and targeting from the platform create discoverable records and are subject to regulatory scrutiny.

We characterise the adversary based on three operational assumptions that reflect the democratised nature of this threat. First, the adversary does not require specialised machine learning expertise or resources to train bespoke classifiers. Instead, they leverage off-the-shelf multimodal LLMs via public APIs (e.g., OpenAI) or local open-source models, making the attack accessible to individuals with only basic technical skills. Second, regarding data access, the adversary is limited to client-side rendered content. They observe the visual and textual components of advertisements exactly as displayed in the user's browser, without requiring privileged access to the ad platform's internal databases. Finally, the adversary operates without access to the user's ground truth private attributes.

\subsection{Attack Vector}

We identify browser extensions that abuse legitimate privileges as the potential primary vector for this attack. This scenario is severe due to its inherent stealth and scalability. Rather than distributing specialised malware, an adversary can opportunistically deploy this attack within the existing ecosystem of widely installed, benign-functioning extensions, such as ad blockers, coupon finders, or page translators. These extensions legitimately require permissions to read web page content to function, providing a perfect cover for data harvesting~\cite{singh2025study}. In contrast, collecting URL histories or full-page logs is more likely to be flagged by extension reviews.

This vector effectively bypasses user and platform scrutiny. Users typically focus their privacy concerns on invisible trackers and cookies while overlooking the semantic signals embedded in visible ad creatives~\cite{yao2017folk}. Similarly, app or plugin store reviews often focus on code safety rather than the privacy implications of data inferred from legitimate content access~\cite{singh2025study}. This creates a regulatory blind spot where a benign-looking extension can silently harvest ad content without triggering security alarms. Furthermore, by leveraging the zero-shot capabilities of LLMs, the adversary can automate this process for scalable profiling. As we demonstrate in Section~\ref{sec:session-level_result}, actionable profiles can be reconstructed from short observation windows, meaning the adversary does not need to maintain long-term persistence to be effective.
Ultimately, this vector offers a distinct strategic advantage: it enables off-platform profiling that evades the platform's own privacy safeguards, such as the removal of sensitive targeting options. The adversary effectively exploits the optimisation logic of the ad platform to build sensitive user profiles at a low marginal cost and without leaving an audit trail.
\section{Methodology}
\label{sec:method}

\begin{figure}[htbp]
    \centering
    \includegraphics[width=0.897\linewidth]{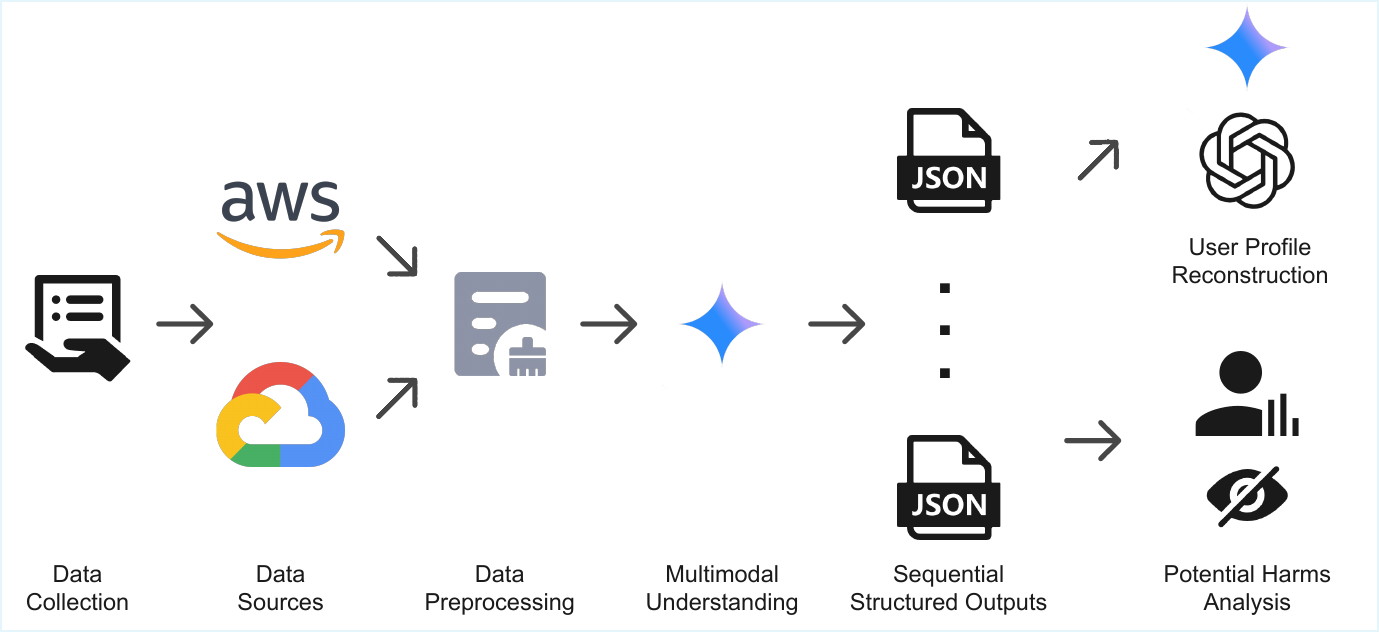}
    \caption{Overall Study Procedure. Begins with data collection from 2000+ Australian Facebook users, then data preprocessing from the data sources. A multimodal LLM generates sequential structured outputs for user profile reconstruction and potential harm analysis.}
    \label{fig:method}
\end{figure}

To empirically evaluate the feasibility of the threat model defined in Section~\ref{sec:threat_model}, we formalise the \textit{User Profile Reconstruction} task. Let $\mathcal{U}$ be a set of users, where each user $u \in \mathcal{U}$ is associated with a ground-truth set of private attributes $\mathcal{A}_u = \{a_{gender}, a_{age}, a_{income}, \dots\}$. The user is exposed to a chronological stream of advertisements $\mathcal{S}_u = (d_1, d_2, \dots, d_n)$, where each ad $d_i$ consists of multimodal content (image $V_i$ and text $T_i$) and metadata.  Our research procedure is illustrated in Figure~\ref{fig:method}. We assume an adversary $\mathcal{M}$, parameterised by an LLM, who seeks to leverage the LLM to perform the mapping function $f: \mathcal{S}_u \to \hat{\mathcal{A}}_u$ that reconstructs the private attributes solely from the ad stream. To achieve scalable and cost-effective inference without model training, we decompose this mapping into a three-stage pipeline: (1) Multimodal Feature Extraction, (2) Session-Level Inference, and (3) Longitudinal User Profiling.

\subsection{Multimodal Ad Understanding}
\label{sec:multimodal_understanding}
To transform raw, multimodal ad data into a structured representation suitable for profiling, we employ an LLM-driven extraction process. Let each advertisement $d_i$ in our dataset consist of visual content $V_i$ (one or more images) and textual metadata $T_i$ (title, body, call-to-action). We utilise Gemini 2.0 Flash~\cite{gemini2_flash_modelcard} as our inference engine $\mathcal{M}$, selected for its balance of multimodal reasoning and cost-efficiency required to process $N > 435,000$ impressions. For each ad, conditioned on an extraction prompt $\mathcal{P}_{\text{extract}}$, the model $\mathcal{M}$ maps the input into a structured feature set $F_i$: 
$F_i = \mathcal{M}(V_i, T_i \mid \mathcal{P}_{\text{extract}}) = \{c_i, \mathcal{D}_i, \mathcal{L}_i, \mathcal{E}_i\}$.

This structured output $F_i$ encapsulates four semantic components: (1) a \textbf{Caption} ($c_i$) summarising the core message and visual elements of the ad; (2) \textbf{Descriptive Categories} ($\mathcal{D}_i$), a set of free-form labels characterising tone, style, and communication strategy; (3) \textbf{IAB Categories} ($\mathcal{L}_i$), a multi-label zero-shot classification into 45 IAB-defined categories~\cite{iab_ad_product_taxonomy}; and (4) \textbf{Key Entities} ($\mathcal{E}_i$), a list of extracted specific brands, products, or locations. This transformation $d_i \to F_i$ extracts high-dimensional multimodal signals into a concise textual representation. We validate the quality of these model-generated features in Appendix~\ref{sec:gemini_quality_eval} and discuss a case of Human–AI disagreement in Appendix~\ref{sec:human-gemini_disagree}.

\subsection{Session-Level User Profile Reconstruction}
\label{sec:method_session}
To capture short-term profiling signals, we segment the continuous user stream $\mathcal{S}_u$ into discrete sessions. Let a session $\mathcal{S}_s$ be a chronologically ordered sequence of $n$ structured ad features observed within a continuous browsing window (segmentation details in Section~\ref{sec:preprocessing}): $\mathcal{S}_s = (F_{s_1}, F_{s_2}, \dots, F_{s_n})$,
where each $F_{s_i}$ corresponds to the structured feature set extracted in the previous stage. We construct a textual representation $T_s$ by concatenating the components of each $F \in \mathcal{S}_s$ in temporal order.

This text-centric transformation is a deliberate methodological choice designed to mitigate two critical limitations of processing raw visual sequences. First, computational feasibility: user sessions vary significantly in length ($n \in [3, 50]$). Processing such long sequences of high-resolution images exceeds the token limits and budget constraints of current API-based models. Second, contextual fidelity: presenting a collage of disparate ad images simultaneously can induce visual-textual alignment ambiguity, where the model struggles to associate specific textual claims with their corresponding visual creatives. We define a session-level inference function that maps the textual representation $T_s$, conditioned on a profiling prompt $\mathcal{P}_{\text{profile session}}$, to a prediction of user's private attributes $\hat{\mathcal{A}}_{s}$ and a reasoning summary $r_s$:
    $(\hat{\mathcal{A}}_{s}, r_s) = \mathcal{M}(T_s \mid \mathcal{P}_{\text{profile session}})$. 
Here, $\hat{\mathcal{A}}_{s}$ contains zero-shot classifications for the six private attributes derived from the current session. The model generates a reasoning summary ($r_s$), a natural language explanation contains the key demographic signals identified within that specific session. This summary serves as the compact information carrier for the subsequent longitudinal analysis.

\subsection{User-Level User Profile Reconstruction}
\label{sec:method_user}
While session-level analysis captures short-term context, a user's full ad exposure history offers a richer and more stable signal. To leverage this, we construct a user-level profile by aggregating the reasoning summaries derived in the previous stage. Let $R_u = (r_{s_1}, r_{s_2}, \dots, r_{s_m})$ denote the chronological sequence of reasoning summaries generated for user $u$ across $m$ sessions. These summaries are concatenated to form a longitudinal narrative, representing a condensed history of the inferred demographic signals over the entire two-year observation period. The final profiling is performed by the model $\mathcal{M}$, conditioned on a synthesis prompt $\mathcal{P}_{\text{profile user}}$, to predict the final attribute set $\hat{\mathcal{A}}_u$:
    $\hat{\mathcal{A}}_u = \mathcal{M}(R_u \mid \mathcal{P}_{\text{profile user}})$.
This hierarchical aggregation allows the adversary to synthesise cumulative evidence (i.e., aggregate semantic signals) and detect temporal patterns (e.g., shifts in interests or trends) from the full reasoning summaries of a user $R_u$. Crucially, this approach remains computationally tractable by processing condensed summaries rather than the prohibitive raw ad stream $\mathcal{S}_u$.
\section{Results}

\begin{table*}[ht!]
\centering
\scriptsize
\begin{tabular}{
    l
    |c@{\hspace{2.2pt}}c|
    c@{\hspace{2.2pt}}c@{\hspace{2.2pt}}c@{\hspace{2.2pt}}c|
    c@{\hspace{2.2pt}}c@{\hspace{2.2pt}}c@{\hspace{2.2pt}}c|
    c@{\hspace{2.2pt}}c|
    c@{\hspace{2.2pt}}c|
    c@{\hspace{2.2pt}}c|
}
\toprule

\multirow{2}{*}{\textbf{Method}}
& \multicolumn{2}{c|}{\textbf{Gender}}
& \multicolumn{4}{c|}{\textbf{Age}}
& \multicolumn{4}{c|}{\textbf{Income}}
& \multicolumn{2}{c|}{\textbf{Education}}
& \multicolumn{2}{c|}{\textbf{Employment}}
& \multicolumn{2}{c}{\textbf{Party}} \\

\cmidrule(lr){2-3}
\cmidrule(lr){4-7}
\cmidrule(lr){8-11}
\cmidrule(lr){12-13}
\cmidrule(lr){14-15}
\cmidrule(lr){16-17}

& Acc (\%) & F1 (\%)
& Acc (\%) & F1 (\%) & MAE & NMAE
& Acc (\%) & F1 (\%) & MAE & NMAE
& Acc (\%) & F1 (\%)
& Acc (\%) & F1 (\%)
& Acc (\%) & F1 (\%) \\
\midrule

Random Guessing
& 50.00 & 50.00
& 14.29 & 14.29 & 2.29 & 0.38
& 8.33  & \textbf{8.33} & 3.97 & 0.36
& 25.00 & \textbf{25.00}
& 20.00 & \textbf{20.00}
& 20.00 & 20.00 \\
\midrule

Random Control
& 58.80 & 58.50
& \textbf{20.56} & \textbf{15.88} & \textbf{1.63} & \textbf{0.27}
& \textbf{10.12}  & 5.04 & \textbf{3.17} & \textbf{0.29}
& \textbf{42.76} & 18.13
& 48.29 & 19.72
& 34.63 & \textbf{21.07} \\
\midrule

Gemini 2.0 Flash
& \textbf{59.13} & \textbf{58.85}
& 20.41 & 15.62 & 1.64 & 0.27
& 9.94 & 4.56 & 3.17 & 0.29
& 42.70 & 18.04
& \textbf{48.38} & 19.92
& \textbf{35.13} & 20.99 \\
\bottomrule
\end{tabular}

\caption{Session-level performance on all 63,864 sessions.}
\label{tab:session_performance}
\end{table*}

\subsection{Dataset}

To empirically evaluate our threat model, we utilise a large-scale, longitudinal dataset sourced from the Australian Ad Observatory, a citizen science project run by the ARC Centre of Excellence for Automated Decision-Making and Society (ADM+S)~\cite{angus2024computational, angus2024australian}. The project recruits volunteer participants from the Australian public to donate data about the advertisements they encounter on Facebook via a custom-built, privacy-preserving browser plugin~\cite{angus2024computational, angus2024australian}. Crucially, this collection mechanism mirrors the exact vantage point of the malicious browser extension described in our threat model (Section~\ref{sec:threat_model}), it passively captures the stream of sponsored posts directly from the user's feed as they browse the website. Participants joined by installing the plugin and completing a detailed demographic questionnaire, which provides the high-fidelity \textit{ground truth} necessary to benchmark adversarial inference accuracy. The collected data for each ad includes the ad creative (image and text), associated metadata, and a link to the de-identified user profile. This methodology provides a unique, user-centric view of the ad ecosystem, capturing the personalised ads that are otherwise inaccessible to public scrutiny via scraping or ad libraries.
The full dataset made available to us comprises over 700,000 ad observations collected between 2021 and 2023 including demographic information from over 2,000 Australian Facebook users. This extensive, real-world dataset allows us to conduct a robust, large-scale study of the potential for user profiling attack by LLMs under realistic conditions.

\subsection{Preprocessing}
\label{sec:preprocessing}

To simulate a realistic attack scenario where an adversary observes users in short discrete time windows, we must segment the continuous stream of ad exposures into meaningful units of analysis. We use a data-driven, principled way to define what counts as a ``session'' in our dataset, instead of arbitrarily choosing a cutoff. Users' ad exposures arrive as a continuous, non-uniform timeline of events over days and weeks. To segment each user's ad viewing history into coherent sessions, we identify a robust threshold for the maximum time interval that can elapse between consecutive ads such that they still belong to the same session.

We compute a session threshold \(\theta\) by applying a kernel density estimator (KDE) over the log-transformed inter-ad time intervals. This process yields a global threshold of \(\theta = 389\) seconds (refer to Appendix~\ref{sec:appendix_preprocessing} for the detailed preprocessing details). Thus, two consecutive ads \(t_i, t_{i+1}\) are considered to belong to different sessions if \(t_{i+1} - t_i > \theta\). This data-driven approach allows us to transform the continuous variable of time gaps into a principled categorical distinction (same session vs. new session) based on real user behavioural patterns observed in the data. Following temporal segmentation, we apply additional filtering steps to ensure data quality and modelling tractability: (1) \textbf{Source filtering}: We retain only ad impressions sourced from Facebook, excluding other platforms and modalities such as video-only content. (2) \textbf{Session length bounds}: We discard sessions with fewer than 3 or more than 50 ad impressions to eliminate sparse sessions and abnormally long sequences. This range aligns with recent studies on LLM-based user profiling and prediction on sequential data~\cite{feng2025agentmove,wongso2024genup}. (3) \textbf{User filtering}: We exclude users who contributed fewer than 3 ad sessions, as such short histories are insufficient for meaningful profiling.

After filtering, the dataset contains: \( 891 \) unique users, \(63{,}864\) total ad sessions and \( 435{,}314 \) total ad impressions. To prepare the ad text for subsequent processing, we remove all HTML elements and markup to produce clean plain-text. The demographic composition of the final filtered user cohort is detailed in Table~\ref{tab:dataset}. This table provides the operational definitions for all demographic categories, such as age ranges (older or younger), income brackets and education levels (higher or lower), used throughout our analyses. All monetary values, including income brackets, are reported in Australian Dollars (AUD). The distribution of ad content across IAB categories, which forms a core feature for our subsequent analyses, is visualised in Figure~\ref{fig:iab_category}. These distributions provide the foundational context for our user profile reconstruction experiments.

\subsection{User Profile Reconstruction}

\begin{table*}[h]
\centering
\scriptsize
\begin{tabular}{
    l
    |c@{\hspace{1.7pt}}c|
    c@{\hspace{1.7pt}}c@{\hspace{1.7pt}}c@{\hspace{1.7pt}}c|
    c@{\hspace{1.7pt}}c@{\hspace{1.7pt}}c@{\hspace{1.7pt}}c|
    c@{\hspace{1.7pt}}c|
    c@{\hspace{1.7pt}}c|
    c@{\hspace{1.7pt}}c|
}
\toprule

\multirow{2}{*}{\textbf{Methods}}
& \multicolumn{2}{c|}{\textbf{Gender}}
& \multicolumn{4}{c|}{\textbf{Age}}
& \multicolumn{4}{c|}{\textbf{Income}}
& \multicolumn{2}{c|}{\textbf{Education}}
& \multicolumn{2}{c|}{\textbf{Employment}}
& \multicolumn{2}{c}{\textbf{Party}} \\

\cmidrule(lr){2-3}
\cmidrule(lr){4-7}
\cmidrule(lr){8-11}
\cmidrule(lr){12-13}
\cmidrule(lr){14-15}
\cmidrule(lr){16-17}

& Acc (\%) & F1 (\%)
& Acc (\%) & F1 (\%) & MAE & NMAE
& Acc (\%) & F1 (\%) & MAE & NMAE
& Acc (\%) & F1 (\%)
& Acc (\%) & F1 (\%)
& Acc (\%) & F1 (\%) \\
\midrule

Random guessing
& 50.00 & 50.00
& 14.29 & 14.29 & 2.29 & 0.38
& 8.33  & 8.33  & 3.97 & 0.36
& 25.00 & 25.00
& 20.00 & 20.00
& 20.00 & 20.00 \\
\midrule

Human
& 73.67\textsubscript{\tiny $\pm$5.13} & 73.05\textsubscript{\tiny $\pm$4.95}
& 21.83\textsubscript{\tiny $\pm$5.34} & 15.64\textsubscript{\tiny $\pm$4.34} & 1.58\textsubscript{\tiny $\pm$0.09} & 0.26\textsubscript{\tiny $\pm$0.01}
& 8.5\textsubscript{\tiny $\pm$2.74}  & 6.94\textsubscript{\tiny $\pm$2.35} & 3.33\textsubscript{\tiny $\pm$0.31} & 0.30\textsubscript{\tiny $\pm$0.03}
& 33.67\textsubscript{\tiny $\pm$5.43} & 25.10\textsubscript{\tiny $\pm$6.41}
& 37.83\textsubscript{\tiny $\pm$5.78} & 20.42\textsubscript{\tiny $\pm$3.95}
& 25.00\textsubscript{\tiny $\pm$6.10} & 16.68\textsubscript{\tiny $\pm$4.54} \\
\midrule

GPT 4o
& 60.00 & 59.86
& \textbf{36.00} & \textbf{27.56} & 1.26 & 0.21
& \textbf{14.00} & \textbf{11.36} & \textbf{2.66} & \textbf{0.24}
& 46.00 & 25.29
& 49.00 & 22.87
& 26.00 & 16.15 \\
\midrule

Gemini 2.0 Flash
& 60.00 & 59.86
& 30.00 & 20.03 & 1.33 & 0.22
& 6.00  & 2.31  & 2.84 & 0.26
& 47.00 & 26.10
& 51.00 & 23.31
& \textbf{32.00} & 15.24 \\
\midrule

Gemini 2.5 Flash
& 64.00 & 63.99
& 30.00 & 26.31 & 1.24 & 0.21
& 7.00  & 3.30  & 3.04 & 0.28
& 43.00 & 24.14
& 49.00 & 22.90
& 25.00 & 13.73 \\
\midrule

\rowcolor{gray!20}
\multicolumn{17}{l}{\emph{Thinking Models}} \\
\midrule

GPT 5 Mini
& 69.00 & 68.97
& 29.00 & 19.72 & 1.29 & 0.22
& 9.00  & 6.60  & 2.9 & 0.26
& 43.00 & 23.59
& 49.00 & 22.62
& 26.00 & 15.81 \\
\midrule

GPT 5
& \textbf{76.00} & \textbf{75.65}
& 31.00 & 24.82 & 1.23 & 0.21
& 11.00 & 5.91  & 2.81 & 0.26
& 46.00 & 27.12
& 51.00 & 24.11
& 25.00 & 14.37 \\
\midrule

Gemini 2.5 Pro
& 75.00 & 74.80
& 30.00 & 26.45 & \textbf{1.15} & \textbf{0.19}
& 8.00  & 5.80  & 3.02 & 0.27
& \textbf{51.00} & \textbf{30.54}
& \textbf{53.00} & \textbf{26.33}
& 31.00 & \textbf{21.72} \\

\bottomrule
\end{tabular}

\caption{Comparison of Human and random guessing against LLMs on 100 sampled human-evaluated sessions.}
\label{tab:human_performance}
\end{table*}

As detailed in Section~\ref{sec:intro} and~\ref{sec:related}, algorithmic optimisation inherently introduces demographic skews in ad delivery. Our preliminary examination of specific categories within our dataset, namely \textit{Education and Careers}, \textit{Gambling}, \textit{Alcohol}, and \textit{Politics}, confirms that these correlations persist locally. Specifically, we observed heightened exposure of gambling content to socioeconomically vulnerable groups, along with distinct age and gender stratifications in alcohol and political advertising. This validates the premise of our threat model: an ad stream serves as a rich digital footprint containing latent private attribute signals.
To quantify the extent to which these signals can be exploited, we deploy the adversarial inference pipeline defined in Section~\ref{sec:method}. Treating the sequence of advertisements as the sole input, we tasked the LLM with performing zero-shot classification of private attributes. We structure our evaluation in two stages: first assessing the feasibility of inference from short observation windows (\textit{Session-Level}), and then aggregating these insights to construct longitudinal profiles (\textit{User-Level}). In both stages, we benchmark performance against strong census-based priors and human evaluators to rigorously characterise the privacy risk.

\textbf{Experimental Setup.} We evaluate performance using Accuracy and Macro F1-score, ensuring robustness under class imbalance. For ordinal attributes such as \textit{Age} and \textit{Income}, where exact classification is overly punitive and ``near misses'' retain profiling utility, we additionally report Mean Absolute Error (MAE). This measures the average index distance between the predicted and ground-truth bins. To allow for comparison across attributes with differing granularities (e.g., 7 age brackets vs. 12 income brackets), we also report Normalised MAE (NMAE). Complete model hyperparameters and configurations are detailed in Appendix~\ref{sec:param_of_models}.

\subsubsection{Session-Level Reconstruction}
\label{sec:session-level_result}

We first conduct our analysis at the most granular level: \textbf{the user session}, defined as the sequence of ads a user is exposed to within a single, continuous period of platform usage. This session-level analysis tests the LLM's inference capabilities on short, contextually coherent ad sequences. To ensure data quality and meaningful classification, we excluded responses where users selected ``prefer not to say'' for any given attribute. We also excluded the ``other'' category for gender due to its small sample size (N=6), which is insufficient for a reliable model evaluation. We then tasked Gemini 2.0 Flash, a state-of-the-art multimodal LLM, with zero-shot classification for six demographic attributes across the remaining 63,864 user sessions from our dataset.

\begin{figure}[htbp]
    \centering
    \includegraphics[width=0.882\linewidth]{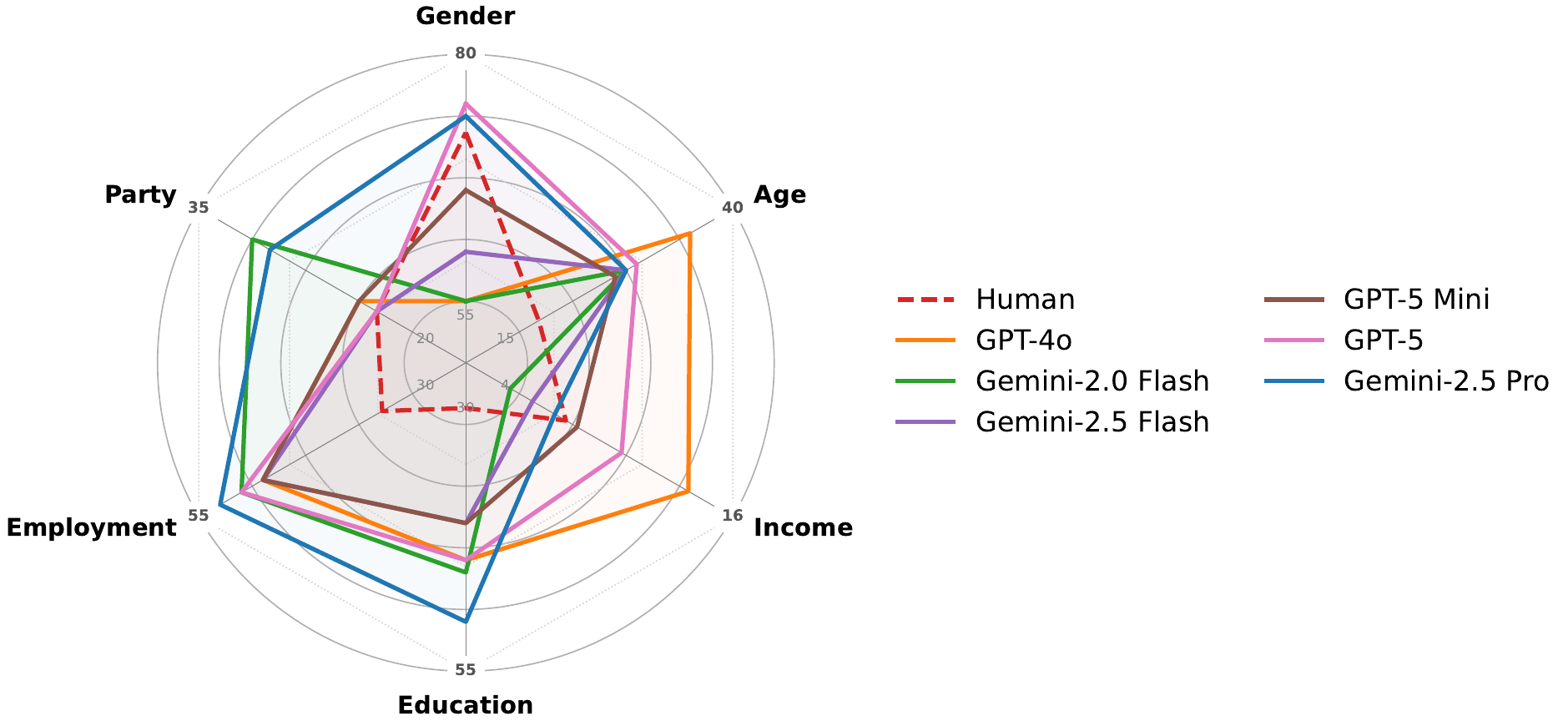}
    \caption{Accuracies of 6 state-of-the-art LLMs and human across different demographic categories on 100 sampled human-evaluated sessions.}
    \label{fig:radar}
\end{figure}

\textbf{Performance on Full Dataset.} 
Table~\ref{tab:session_performance} presents the model's performance benchmarked against a random guessing baseline and a ``Random Control'' group (where ad order within a session is shuffled).
For categorical attributes, Gemini 2.0 Flash consistently outperforms the random baseline. For instance, the model achieves 59.13\% accuracy for \textit{Gender}, and substantial gains are observed for \textit{Education} (42.70\%) and \textit{Employment} (48.38\%), more than doubling the random guess performance. Similarly, for \textit{Party} preference, the model achieves 35.13\% compared to the 20.00\% baseline, showing that political orientation is discernible even within short browsing sessions. This indicates that ad sequences, even at the session level, contain sufficient demographic signals for the LLM to make inferences better than chance. For ordinal attributes (\textit{Age} and \textit{Income}). While the F1-scores for \textit{Age} (15.62\%) and \textit{Income} (4.56\%) appear low due to the difficulty of pinpointing exact brackets, the MAE metrics reveal a deeper insight. The model achieves an MAE of 1.64 and 3.17 (NMAE 0.27 and 0.29) for \textit{Age} and \textit{Income}, significantly lower than the random baseline of 2.29 and 3.97. This implies that the model's errors are not random noise, rather, predictions are typically ``near misses'' falling within 1-2 buckets of the truth. This suggests that ad streams contain a robust signal for a user's approximate life stage and economic standing, which the model effectively decodes even if exact precision remains challenging.

\textbf{Impact of Temporal Order (Random Control).}
To isolate the value of sequential information at the session level, we compare the standard model against the Random Control group in Table~\ref{tab:session_performance}. The performance difference is negligible across all metrics. For example, \textit{Party} preference accuracy shifts marginally from 34.63\% (Control) to 35.13\% (Sequential). This finding confirms that at the granular level of a single, short user session, the specific chronological order of ads provides little additional signal. The model's inferences appear to be based primarily on the aggregate semantic content present in the session, rather than their temporal arrangement.

\begin{table*}[h]
\centering
\scriptsize
\begin{tabular}{
    l
    |c@{\hspace{1.7pt}}c|
    c@{\hspace{1.7pt}}c@{\hspace{1.7pt}}c@{\hspace{1.7pt}}c|
    c@{\hspace{1.7pt}}c@{\hspace{1.7pt}}c@{\hspace{1.7pt}}c|
    c@{\hspace{1.7pt}}c|
    c@{\hspace{1.7pt}}c|
    c@{\hspace{1.7pt}}c|
}
\toprule

\multirow{2}{*}{\textbf{Method}}
& \multicolumn{2}{c|}{\textbf{Gender}}
& \multicolumn{4}{c|}{\textbf{Age}}
& \multicolumn{4}{c|}{\textbf{Income}}
& \multicolumn{2}{c|}{\textbf{Education}}
& \multicolumn{2}{c|}{\textbf{Employment}}
& \multicolumn{2}{c}{\textbf{Party}} \\

\cmidrule(lr){2-3}
\cmidrule(lr){4-7}
\cmidrule(lr){8-11}
\cmidrule(lr){12-13}
\cmidrule(lr){14-15}
\cmidrule(lr){16-17}

& Acc (\%) & F1 (\%)
& Acc (\%) & F1 (\%) & MAE & NMAE
& Acc (\%) & F1 (\%) & MAE & NMAE
& Acc (\%) & F1 (\%)
& Acc (\%) & F1 (\%)
& Acc (\%) & F1 (\%) \\
\midrule

Prior-Mode
& 36.47 & 26.72
& 20.84 & 4.93 & 2.04 & 0.34
& 9.25  & 1.54 & 3.07 & 0.28
& 21.81 & 11.94
& 56.71 & 24.13
& 3.14 & 1.22 \\
\midrule

Prior-Sampling
& 49.89\textsubscript{\tiny $\pm$1.68} & 49.03\textsubscript{\tiny $\pm$1.68}
& 14.85\textsubscript{\tiny $\pm$1.22} & 14.01\textsubscript{\tiny $\pm$1.18} & 2.11\textsubscript{\tiny $\pm$0.05} & 0.35\textsubscript{\tiny $\pm$0.01}
& 8.32\textsubscript{\tiny $\pm$0.96}  & \textbf{8.03\textsubscript{\tiny $\pm$0.95}}  & 4.06\textsubscript{\tiny $\pm$0.10} & 0.37\textsubscript{\tiny $\pm$0.01}
& 26.63\textsubscript{\tiny $\pm$1.24} & 24.20\textsubscript{\tiny $\pm$1.38}
& 46.90\textsubscript{\tiny $\pm$1.79} & 32.77\textsubscript{\tiny $\pm$1.77}
& 19.85\textsubscript{\tiny $\pm$1.25} & 16.20\textsubscript{\tiny $\pm$1.13} \\
\midrule

Random Control
& 70.39 & 70.38
& 38.00 & 32.02 & 0.90 & 0.15
& \textbf{10.27} & 6.31 & \textbf{2.73} & \textbf{0.25}
& 43.48 & 33.60
& 61.36 & 33.59
& 39.92 & 23.81 \\
\midrule

Gemini 2.0 Flash
& 74.88 & 74.87
& \textbf{41.18} & \textbf{34.57} & \textbf{0.85} & \textbf{0.14}
& 9.11 & 5.46 & 2.73 & 0.25
& \textbf{44.87} & 34.13
& 61.95 & 34.66
& 39.91 & 23.90 \\
\midrule

Gemini 2.0 Flash$^\dag$
& \textbf{76.38} & \textbf{76.35}
& 40.84 & 34.39 & 0.95 & 0.16
& 8.99 & 5.76 & 2.77 & 0.25
& 44.41 & \textbf{34.37}
& \textbf{62.46} & \textbf{35.63}
& \textbf{41.60} & \textbf{25.82} \\

\bottomrule
\end{tabular}
\caption{User-level performance on 891 users. $\dag$ denotes the LLM augmented with an Australian-specific context prompt.}
\label{tab:user_performance}
\end{table*}

\textbf{Human vs. AI Evaluation.}
To contextualise these LLMs' capabilities, we conducted a human evaluation on randomly sampled 100 user sessions, stratified to match the demographic distribution of our dataset as detailed in Table~\ref{tab:dataset}. We recruited six human annotators (2 female, 4 male) with diverse professional backgrounds, including legal and privacy expertise, computer science, and human resources.  Each annotator was tasked with inferring the six demographic attributes for each of the 100 sessions, based solely on the same ad content shown to the LLM. To best capture intuitive human inference in a realistic and tractable manner, annotators were shown the raw, human-readable visual images of the ads within each session, as processing the full structured textual captions, categories, key entities from all ads in 100 sessions would be cognitively overwhelming. In contrast, the LLM received the structured textual input generated by our multimodal understanding pipeline (as described in Section~\ref{sec:multimodal_understanding}) as in our session-level analysis. This design allows for a valid comparison between intuitive human inference and systematic algorithmic inference. The results are detailed in Table~\ref{tab:human_performance} and visualised in Figure~\ref{fig:radar}.

The comparison reveals a significant evolution in the capabilities regarding \textit{Gender} classification. While this task relies heavily on decoding subtle, socially-cued signals, a domain where human intuition typically excels (73.67\% accuracy), state-of-the-art LLMs now demonstrate remarkable proficiency that rivals or even exceeds human performance (GPT-5 at 76\% and Gemini 2.5 Pro at 75\%). This indicates that modern multimodal models have reached a level of social perception comparable to humans, effectively bridging the gap in interpreting nuanced demographic cues from visual content. Furthermore, for complex socio-economic and political attributes, LLMs consistently matched or surpassed human capabilities. For \textit{Education} and \textit{Employment}, the best-performing model (Gemini 2.5 Pro) achieved significantly higher accuracies (51\% and 53\%) compared to the human average (33.67\% and 37.83\%). A similar trend was observed for \textit{Party} preference, where the model reached 32\% accuracy against the human baseline of 25\%. This indicates that LLMs are more effective at identifying non-obvious correlations between commercial content and users' professional backgrounds or ideological leanings. Crucially, the MAE analysis for \textit{Age} highlights the superior calibration of LLMs. While humans achieved an MAE of 1.58, the best LLM (Gemini 2.5 Pro) achieved 1.15. This confirms that even when the model fails to predict the exact age bracket, its estimates are semantically closer to the ground truth than human intuition. Both humans and LLMs struggled with \textit{Income}, reinforcing that this attribute is less reliably encoded in short-term ad content.

In summary, our \textbf{session-level} analysis provides empirical evidence that LLMs can extract private attribute signals from short observation windows, achieving performance that rivals or even exceeds human capabilities. The models demonstrate a ``directional correctness'' (low MAE) for ordinal attributes and significantly outperform human annotators on complex traits like employment and education. This feasibility of short-term profiling motivates our subsequent analysis at the user level, where aggregating information across multiple sessions may amplify these signals.

\subsubsection{User-Level Reconstruction}

Building on the hierarchical aggregation pipeline defined in Section~\ref{sec:method_user}, we reconstruct longitudinal user profiles by synthesizing reasoning summaries across all sessions. This holistic approach allows the model to leverage cumulative evidence and temporal shifts in ad exposure over the user's entire history.
For this more rigorous user-level evaluation, we introduce two stronger baseline models derived from Australian census data~\cite{ABS2021Census}, which reflect the prior demographic distribution of the population. The Prior-Mode baseline always predicts the majority class for each attribute. The Prior-Sampling baseline provides a more robust chance-level comparison by making predictions through random sampling from the census distribution; we report the mean and standard deviation of its performance over 1,000 independent runs in Table~\ref{tab:user_performance}. To ensure a fair comparison, we harmonised our dataset with the census data. For all attributes, we excluded responses where users selected ``prefer not to say.'' We excluded gender-diverse respondents from this analysis because the census prior reports only male and female categories. For education, we aligned with the higher education census distribution by merging all non-degree categories in our data into a single ``No Degree'' class. Census data for individual income, reported weekly, was annualised and aligned perfectly with our income brackets. For employment, we merged our two ``unemployed'' categories to match the census, which does not distinguish between looking for work or not; we also excluded ``retired'' individuals from this specific attribute's evaluation, as the census employment data pertains only to the labour force. Detailed demographic statistics are based on the Australian Census~\cite{ABS2021Census} and the Australian Federal Election Study~\cite{cameron20222022}. We evaluate three versions of LLM: the standard Gemini 2.0 flash model, an augmented version, Gemini$^\dag$, which was provided with a prompt about the Australian cultural context and a ``Random Control'' group (where ad session order within a user is shuffled). This augmentation is designed to create a more equitable comparison with our census-derived baselines. Since the Prior-Sampling baseline inherently incorporates ``Australian context'' by drawing from national demographic data, providing the LLM with similar high-level contextual information ensures that our evaluation more accurately assesses the model's ability to infer information from ad content, rather than simply penalising it for a lack of geographical and cultural knowledge.

The \textbf{user-level} reconstruction performance is presented in Table~\ref{tab:user_performance}. The results show a marked improvement over the session-level analysis and demonstrate the LLM's strong capability to reconstruct user profiles. Gemini significantly outperforms both the Prior-Mode and Prior-Sampling baselines across nearly all attributes in both accuracy and F1-score. The most striking performance is in \textit{Gender} reconstruction, where Gemini$^\dag$ achieves 76.38\% accuracy and a 76.35\% F1-score, far exceeding the baselines and indicating a very strong and reliable signal in the ad streams. Substantial gains are also evident for attributes that were challenging at the session level. For \textit{Education}, the F1-score jumps to 34.37\%, nearly tripling the baseline performance. The model's performance on \textit{Employment} is also noteworthy. While the Prior-Mode baseline achieves a high accuracy of 56.71\% due to a large majority class (employed full-time), its F1-score is low (24.13\%). Gemini$^\dag$, in contrast, achieves a higher accuracy (62.46\%) and a significantly better F1-score (35.63\%), demonstrating a superior ability to correctly identify users in non-majority employment categories. For \textit{Party} preference, the LLM's accuracy (41.60\%) and F1-score (25.82\%) are dramatically better than the baselines, suggesting that political orientation is encoded in ad content. For ordinal attributes, \textit{Age}, Gemini's accuracy (41.18\%) is double that of the Prior-Mode baseline (20.84\%), but the MAE provides a deeper insight: the model achieves an MAE of 0.85 (NMAE 0.14), which is significantly lower than the Prior-Mode (2.04) and Prior-Sampling (2.11). An MAE below 1.0 indicates that on average, the model's predictions are less than one age bracket away from the ground truth. This confirms that the model successfully identifies the user's approximate life stage with high ``directional correctness.'' A similar pattern is observed for \textit{Income}. While the absolute accuracy (9.11\%) remains low and comparable to baselines, the MAE of 2.73 is substantially better than the Prior-Sampling baseline (4.06). This reduction in error magnitude suggests that while pinpointing the exact income bracket remains difficult, the ad content contains a coarse but discernible signal of a user's general economic standing, allowing the model to place users in brackets semantically closer to the truth than random chance.

To isolate sequential information, we compare the standard model against a ``Random Control'' group where the order of ads within sessions was shuffled. The results indicate that temporal order contributes differently across attributes. For \textit{Age} and \textit{Gender}, preserving the chronological sequence yields gains (e.g., Age accuracy improves from 38.00\% to 41.18\%, and MAE decreases to 0.85), suggesting that temporal evolution in ad delivery captures signals related to life stages or identity stability. However, for socio-economic attributes like \textit{Income}, \textit{Education}, and \textit{Employment}, performance remains comparable between the sequential and shuffled inputs. This implies that for these traits, the cumulative semantic content: the specific collection of ads a user sees, serves as the primary predictive signal, regardless of the specific viewing order.

In summary, our analysis reveals a clear privacy risk. At the \textbf{session level}, LLMs can extract actionable signals from short observation windows; at the \textbf{user level}, aggregating these signals significantly amplifies profiling precision. The strong performance of the Random Control group and non-augmented version Gemini demonstrates that the mere accumulation of ad exposures creates a high-fidelity digital fingerprint even without temporal structure and cultural context. This confirms that passive ad streams are not just ephemeral noise but a leakage vector that allows LLMs to reverse-engineer private attributes with alarming accuracy.
\section{Discussion}

\textbf{The Shift to Passive Privacy Risk.}
Our findings mark a shift in digital privacy understanding. Our LLM-based reconstruction, which outperforms census priors and matches human intuition, confirms that generative AI can systematically decode granular demographic signals encoded in ad-delivery algorithms. Furthermore, our analysis of ordinal attributes (\textit{Age}, \textit{Income}) reveals that when exact predictions fail, they remain ``directionally correct'' (low MAE), placing users in semantically accurate categories (e.g., ``young'', ``low-income''). The democratisation of multimodal LLMs transforms this from a theoretical risk into an immediate operational threat. Unlike traditional profiling that required collecting massive labeled datasets to train bespoke classifiers~\cite{bi2013inferring,kosinski2013private,schwartz2013personality,hinds2018demographic, tricomi2024spotify}, our zero-shot approach demonstrates that adversaries can now leverage off-the-shelf models to perform scalable profiling with minimal technical expertise. While prior research demonstrated that LLMs can infer attributes from user content such as Reddit posts~\cite{staab24beyond, tomekcce2024private} or personal photo albums~\cite{liu2025eye}, our work reveals that \textit{passive} consumption of algorithmic content constitutes a comparable leakage vector. This capability is particularly dangerous within the browser extension ecosystem~\cite{singh2025study}, where benign-looking tools (e.g., ad blockers) possess legitimate permissions to read page content. An adversary can thus deploy a ``silent'' profiler that bypasses platform targeting restrictions (e.g., on political affiliation) by reverse-engineering the platform's own optimisation logic locally on the user's device. This vector exploits a critical gap in user understanding: users typically focus privacy concerns on invisible trackers and cookies, rarely perceiving the visible ad creative itself as a leakage vector~\cite{yao2017folk}. This creates a privacy dilemma: users can choose not to post content, but they cannot easily opt out of the ad ecosystem, and they are often unaware that even passive exposure can be harvested to reconstruct their identity.  Taken together, these findings suggest that current privacy policy may underestimate risks by focusing on exact identification rather than semantic profiling.

\textbf{Limitations.}
We note several limitations. First, our dataset is drawn from a self-selected cohort of Australian Facebook users via a desktop browser plugin. This sample may not fully generalize to mobile-only populations or other cultural and regulatory contexts~\cite{kaushik2024cross}. Second, our analysis is limited to Facebook; other ecosystems like TikTok or open programmatic web advertising involve different ad formats and delivery logics that may yield different leakage patterns. Finally, our reconstruction pipeline separates visual captioning from reasoning to manage token costs. While effective, a fully end-to-end multimodal model might capture even subtler visual cues, potentially yielding higher inference accuracy and greater privacy risk than reported here. Future work should extend this framework to cross-platform environments and explore adversarial defenses against LLM-based profiling.
\section{Conclusion}

This work reveals a critical blind spot in Web privacy: the latent leakage of user private attributes through passive exposure to algorithmic advertising. By leveraging multimodal LLMs as adversarial inference engines, we demonstrate that ad streams serve as high-fidelity digital footprints, enabling the accurate reconstruction of private attributes, often surpassing human social perception. Our findings confirm that this threat is operational and scalable: actionable profiles can be derived from short observation windows without long-term tracking, and ``near-miss'' predictions for age and income yield semantically invasive insights. As off-the-shelf AI tools democratise this capability, they increase the risk that benign browser extensions could be repurposed to circumvent platform targeting restrictions. Addressing this risk requires moving beyond current regulatory frameworks to recognise and govern the latent semantic signals embedded in the content users passively consume.

\begin{acks}
This research includes computations using the Wolfpack computational cluster, supported by the School of Computer Science and Engineering at UNSW Sydney. We acknowledge support from the Australian Research Council (ARC) Centre of Excellence for Automated Decision-Making and Society (ADM+S; CE200100005), which hosts the Australian Ad Observatory project, and thank the Ad Observatory team for their work and support. We thank Abdul Obeid for suggesting the idea of using a KDE-based approach for defining session thresholds. We further thank our annotators for their hard work, which greatly contributed to the human evaluation. 
\end{acks}

\bibliographystyle{ACM-Reference-Format}
\bibliography{sample-base}

@String{Computing = "Computing" }

@String{Computer = "{IEEE} Computer" }

@inproceedings{AI_Driven_Online_Advertising_Market_Design_Generative_AI_and_Ethics,
author = {He, Fengxiang and Du, Mengnan and Filos-Ratsikas, Aris and Cheng, Lu and Song, Qingquan and Lin, Min and Vines, John},
title = {AI Driven Online Advertising: Market Design, Generative AI, and Ethics},
year = {2024},
isbn = {9798400701726},
publisher = {Association for Computing Machinery},
address = {New York, NY, USA},
doi = {10.1145/3589335.3641295},
booktitle = {Companion Proceedings of the ACM Web Conference 2024},
pages = {1407–1409},
numpages = {3},
location = {Singapore, Singapore},
series = {WWW '24}
}

@inproceedings{Computational_Advertising_Recent_Advances,
author = {Pan, Junwei and Zhang, Zhilin and Zhu, Han and Xu, Jian and Jiang, Jie and Zheng, Bo},
title = {Computational Advertising: Recent Advances},
year = {2025},
isbn = {9798400713316},
publisher = {Association for Computing Machinery},
address = {New York, NY, USA},
doi = {10.1145/3701716.3715872},
pages = {37–40},
numpages = {4},
location = {Sydney NSW, Australia},
series = {WWW '25}
}

@article{angus2024computational,
  title     = {Computational Methods for Improving the Observability of Platform-Based Advertising},
  author    = {Angus, Daniel and Hayden, Lauren and Obeid, Abdul Karim and Tan, Xue Ying and Carah, Nicholas and Burgess, Jean and Parker, Christine and others},
  journal   = {Journal of Advertising},
  volume    = {53},
  number    = {5},
  pages     = {661--680},
  year      = {2024},
  doi       = {10.1080/00913367.2024.2394156},
  url       = {https://doi.org/10.1080/00913367.2024.2394156}
}

@article{zhou2024language,
  title={Language-based user profiles for recommendation},
  author={Zhou, Joyce and Dai, Yijia and Joachims, Thorsten},
  journal={arXiv preprint arXiv:2402.15623},
  year={2024}
}

@inbook{wongso2024genup,
author = {Wongso, Wilson and Xue, Hao and Salim, Flora},
title = {GenUP: Generative User Profilers as In-Context Learners for Next POI Recommender Systems},
year = {2025},
isbn = {9798400720864},
publisher = {Association for Computing Machinery},
address = {New York, NY, USA},
booktitle = {Proceedings of the 33rd ACM International Conference on Advances in Geographic Information Systems},
pages = {436–439},
numpages = {4}
}

@inproceedings{ramos-etal-2024-transparent,
    title = "Transparent and Scrutable Recommendations Using Natural Language User Profiles",
    author = "Ramos, Jerome  and
      Rahmani, Hossein A.  and
      Wang, Xi  and
      Fu, Xiao  and
      Lipani, Aldo",
    editor = "Ku, Lun-Wei  and
      Martins, Andre  and
      Srikumar, Vivek",
    booktitle = "Proceedings of the 62nd Annual Meeting of the Association for Computational Linguistics (Volume 1: Long Papers)",
    month = aug,
    year = "2024",
    address = "Bangkok, Thailand",
    publisher = "Association for Computational Linguistics",
    url = "https://aclanthology.org/2024.acl-long.753/",
    doi = "10.18653/v1/2024.acl-long.753",
    pages = "13971--13984"
}

@misc{angus2024australian,
  title={The Australian Ad Observatory technical and data report},
  author={Angus, Daniel and Obeid, Abdul and Burgess, Jean and Parker, Christine and Andrejevic, Mark and Bagnara, Julian and Carah, Nicholas and Fordyce, Robbie and Hayden, Lauren and Lewis, Kelly and others},
  year={2024},
  publisher={ARC Centre of Excellence for Automated Decision-Making and Society}
}

@article{ali2019discrimination,
  title={Discrimination through optimization: How Facebook's Ad delivery can lead to biased outcomes},
  author={Ali, Muhammad and Sapiezynski, Piotr and Bogen, Miranda and Korolova, Aleksandra and Mislove, Alan and Rieke, Aaron},
  journal={Proceedings of the ACM on human-computer interaction},
  volume={3},
  number={CSCW},
  pages={1--30},
  year={2019},
  publisher={ACM New York, NY, USA}
}

@inproceedings{balog2019transparent,
  title={Transparent, scrutable and explainable user models for personalized recommendation},
  author={Balog, Krisztian and Radlinski, Filip and Arakelyan, Shushan},
  booktitle={Proceedings of the 42nd international acm sigir conference on research and development in information retrieval},
  pages={265--274},
  year={2019}
}

@inproceedings{radlinski2022natural,
  title={On natural language user profiles for transparent and scrutable recommendation},
  author={Radlinski, Filip and Balog, Krisztian and Diaz, Fernando and Dixon, Lucas and Wedin, Ben},
  booktitle={Proceedings of the 45th international ACM SIGIR conference on research and development in information retrieval},
  pages={2863--2874},
  year={2022}
}

@article{tan2023user,
  title={User modeling in the era of large language models: Current research and future directions},
  author={Tan, Zhaoxuan and Jiang, Meng},
  journal={arXiv preprint arXiv:2312.11518},
  year={2023}
}

@inproceedings{ren2024representation,
  title={Representation learning with large language models for recommendation},
  author={Ren, Xubin and Wei, Wei and Xia, Lianghao and Su, Lixin and Cheng, Suqi and Wang, Junfeng and Yin, Dawei and Huang, Chao},
  booktitle={Proceedings of the ACM Web Conference 2024},
  pages={3464--3475},
  year={2024}
}

@inproceedings{xi2024towards,
  title={Towards open-world recommendation with knowledge augmentation from large language models},
  author={Xi, Yunjia and Liu, Weiwen and Lin, Jianghao and Cai, Xiaoling and Zhu, Hong and Zhu, Jieming and Chen, Bo and Tang, Ruiming and Zhang, Weinan and Yu, Yong},
  booktitle={Proceedings of the 18th ACM Conference on Recommender Systems},
  pages={12--22},
  year={2024}
}

@article{comanici2025gemini,
  title={Gemini 2.5: Pushing the frontier with advanced reasoning, multimodality, long context, and next generation agentic capabilities},
  author={Comanici, Gheorghe and Bieber, Eric and Schaekermann, Mike and Pasupat, Ice and Sachdeva, Noveen and Dhillon, Inderjit and Blistein, Marcel and Ram, Ori and Zhang, Dan and Rosen, Evan and others},
  journal={arXiv preprint arXiv:2507.06261},
  year={2025}
}

@article{hurst2024gpt,
  title={Gpt-4o system card},
  author={Hurst, Aaron and Lerer, Adam and Goucher, Adam P and Perelman, Adam and Ramesh, Aditya and Clark, Aidan and Ostrow, AJ and Welihinda, Akila and Hayes, Alan and Radford, Alec and others},
  journal={arXiv preprint arXiv:2410.21276},
  year={2024}
}

@techreport{anthropic2025claude4,
  title        = {Claude Opus 4 \& Claude Sonnet 4: System Card},
  author       = {{Anthropic}},
  institution  = {Anthropic},
  month        = may,
  year         = {2025},
}

@article{guo2025deepseek,
  title={Deepseek-r1: Incentivizing reasoning capability in llms via reinforcement learning},
  author={Guo, Daya and Yang, Dejian and Zhang, Haowei and Song, Junxiao and Zhang, Ruoyu and Xu, Runxin and Zhu, Qihao and Ma, Shirong and Wang, Peiyi and Bi, Xiao and others},
  journal={arXiv preprint arXiv:2501.12948},
  year={2025}
}

@techreport{openai2025gpt5systemcard,
  title        = {GPT-5 System Card},
  author       = {{OpenAI}},
  institution  = {OpenAI},
  month        = aug,
  day          = 7,
  year         = 2025,
}

@article{jaech2024openai,
  title={Openai o1 system card},
  author={Jaech, Aaron and Kalai, Adam and Lerer, Adam and Richardson, Adam and El-Kishky, Ahmed and Low, Aiden and Helyar, Alec and Madry, Aleksander and Beutel, Alex and Carney, Alex and others},
  journal={arXiv preprint arXiv:2412.16720},
  year={2024}
}

@article{hinds2018demographic,
  title={What demographic attributes do our digital footprints reveal? A systematic review},
  author={Hinds, Joanne and Joinson, Adam N},
  journal={PloS one},
  volume={13},
  number={11},
  pages={e0207112},
  year={2018},
  publisher={Public Library of Science San Francisco, CA USA}
}

@inproceedings{ali2022all,
  title={All things unequal: Measuring disparity of potentially harmful ads on facebook},
  author={Ali, Muhammad and Goetzen, Angelica and Mislove, Alan and Redmiles, Elissa and Sapiezynski, Piotr},
  booktitle={Proceedings of the 2022 workshop on consumer protection},
  year={2022}
}

@inproceedings{ali2023problematic,
  title={Problematic advertising and its disparate exposure on Facebook},
  author={Ali, Muhammad and Goetzen, Angelica and Mislove, Alan and Redmiles, Elissa M and Sapiezynski, Piotr},
  booktitle={32nd USENIX Security Symposium (USENIX Security 23)},
  pages={5665--5682},
  year={2023}
}

@misc{iab_ad_product_taxonomy,
  title        = {{Ad Product Taxonomy}},
  author       = {{IAB Technology Laboratory}},
  howpublished = {IAB Tech Lab website},
  note         = {Last updated: December 11, 2024},
  url          = {https://iabtechlab.com/standards/ad-product-taxonomy/},
}

@techreport{gemini2_flash_modelcard,
  title        = {{Gemini 2.0 Flash Model Card}},
  author       = {{Google DeepMind}},
  institution  = {Google DeepMind},
  month        = jun,
  year         = 2025,
}

@inproceedings{meguellati2024how,
author = {Meguellati, Elyas and Han, Lei and Bernstein, Abraham and Sadiq, Shazia and Demartini, Gianluca},
title = {How Good are LLMs in Generating Personalized Advertisements?},
year = {2024},
isbn = {9798400701726},
publisher = {Association for Computing Machinery},
address = {New York, NY, USA},
booktitle = {Companion Proceedings of the ACM Web Conference 2024},
pages = {826–829},
numpages = {4},
location = {Singapore, Singapore},
series = {WWW '24}
}

@inproceedings{imana2021job,
author = {Imana, Basileal and Korolova, Aleksandra and Heidemann, John},
title = {Auditing for Discrimination in Algorithms Delivering Job Ads},
year = {2021},
isbn = {9781450383127},
publisher = {Association for Computing Machinery},
address = {New York, NY, USA},
doi = {10.1145/3442381.3450077},
pages = {3767–3778},
numpages = {12},
location = {Ljubljana, Slovenia},
series = {WWW '21}
}

@article{
zhang2025personalization,
title={Personalization of Large Language Models: A Survey},
author={Zhehao Zhang and Ryan A. Rossi and Branislav Kveton and Yijia Shao and Diyi Yang and Hamed Zamani and Franck Dernoncourt and Joe Barrow and Tong Yu and Sungchul Kim and Ruiyi Zhang and Jiuxiang Gu and Tyler Derr and Hongjie Chen and Junda Wu and Xiang Chen and Zichao Wang and Subrata Mitra and Nedim Lipka and Nesreen K. Ahmed and Yu Wang},
journal={Transactions on Machine Learning Research},
issn={2835-8856},
year={2025},
}

@misc{openai2025memory,
  title = {Memory and new controls for ChatGPT},
  author = {OpenAI},
  year = {2024},
  howpublished = {\url{https://openai.com/index/memory-and-new-controls-for-chatgpt/}},
}

@article{kosinski2013private,
author = {Michal Kosinski  and David Stillwell  and Thore Graepel },
title = {Private traits and attributes are predictable from digital records of human behavior},
journal = {Proceedings of the National Academy of Sciences},
volume = {110},
number = {15},
pages = {5802-5805},
year = {2013},
doi = {10.1073/pnas.1218772110},
URL = {https://www.pnas.org/doi/abs/10.1073/pnas.1218772110},
eprint = {https://www.pnas.org/doi/pdf/10.1073/pnas.1218772110},
}

@inproceedings{yao2017folk,
author = {Yao, Yaxing and Lo Re, Davide and Wang, Yang},
title = {Folk Models of Online Behavioral Advertising},
year = {2017},
isbn = {9781450343350},
publisher = {Association for Computing Machinery},
address = {New York, NY, USA},
doi = {10.1145/2998181.2998316},
booktitle = {Proceedings of the 2017 ACM Conference on Computer Supported Cooperative Work and Social Computing},
pages = {1957–1969},
numpages = {13},
location = {Portland, Oregon, USA},
series = {CSCW '17}
}

@inproceedings{bi2013inferring,
author = {Bi, Bin and Shokouhi, Milad and Kosinski, Michal and Graepel, Thore},
title = {Inferring the demographics of search users: social data meets search queries},
year = {2013},
isbn = {9781450320351},
publisher = {Association for Computing Machinery},
address = {New York, NY, USA},
doi = {10.1145/2488388.2488401},
booktitle = {Proceedings of the 22nd International Conference on World Wide Web},
pages = {131–140},
numpages = {10},
location = {Rio de Janeiro, Brazil},
series = {WWW '13}
}

@inproceedings{kaushik2024cross,
author = {Kaushik, Smirity and Sharma, Tanusree and Yu, Yaman and Ali, Amna F and Wang, Yang and Zou, Yixin},
title = {Cross-Country Examination of People’s Experience with Targeted Advertising on Social Media},
year = {2024},
isbn = {9798400703317},
publisher = {Association for Computing Machinery},
address = {New York, NY, USA},
doi = {10.1145/3613905.3650780},
booktitle = {Extended Abstracts of the CHI Conference on Human Factors in Computing Systems},
articleno = {91},
numpages = {10},
location = {Honolulu, HI, USA},
series = {CHI EA '24}
}

@misc{GDPR2016,
  title        = {Regulation (EU) 2016/679 of the European Parliament and of the Council (General Data Protection Regulation)},
  author       = {{European Parliament and Council of the European Union}},
  year         = {2016},
  howpublished = {Official Journal of the European Union, L119},
  note         = {OJ L 119, 4.5.2016}
}

@article{singh2025study,
  title={A study on malicious browser extensions in 2025},
  author={Singh, Shreya and Varshney, Gaurav and Singh, Tarun Kumar and Mishra, Vidhi and Verma, Khushi},
  journal={arXiv preprint arXiv:2503.04292},
  year={2025}
}

@article{cameron20222022,
  title={The 2022 Australian federal election: Results from the Australian election study},
  author={Cameron, Sarah and McAllister, Ian and Jackman, Simon and Sheppard, Jill},
  year={2022},
  publisher={The Australian Election Study}
}

@misc{ABS2021Census,
  author       = {{Australian Bureau of Statistics}},
  title        = {{Population: Census, 2021}},
  year         = {2021},
  publisher    = {{Australian Bureau of Statistics}},
  address      = {Canberra},
  howpublished = {\url{https://www.abs.gov.au/statistics/people/population/population-census/2021}},
  note         = {Accessed on August 2025}
}

@article{schwartz2013personality,
  title={Personality, gender, and age in the language of social media: The open-vocabulary approach},
  author={Schwartz, H Andrew and Eichstaedt, Johannes C and Kern, Margaret L and Dziurzynski, Lukasz and Ramones, Stephanie M and Agrawal, Megha and Shah, Achal and Kosinski, Michal and Stillwell, David and Seligman, Martin EP and others},
  journal={PloS one},
  volume={8},
  number={9},
  pages={e73791},
  year={2013},
  publisher={Public Library of Science}
}

@inproceedings{staab24beyond,
    title={Beyond Memorization: Violating Privacy via Inference with Large Language Models},
    author={Robin Staab and Mark Vero and Mislav Balunović and Martin Vechev},
    booktitle={The Twelfth International Conference on Learning Representations},
    year={2024},
}

@inproceedings{liu2025eye,
author = {Liu, Feiran and Zhang, Yuzhe and Huang, Xinyi and Peng, Yinan and Li, Xinfeng and Wang, Lixu and Shen, Yutong and Duan, Ranjie and Qin, Simeng and Jia, Xiaojun and Wen, Qingsong and Dong, Wei},
title = {The Eye of Sherlock Holmes: Uncovering User Private Attribute Profiling via Vision-Language Model Agentic Framework},
year = {2025},
isbn = {9798400720352},
publisher = {Association for Computing Machinery},
address = {New York, NY, USA},
doi = {10.1145/3746027.3755643},
booktitle = {Proceedings of the 33rd ACM International Conference on Multimedia},
pages = {4875–4883},
numpages = {9},
location = {Dublin, Ireland},
series = {MM '25}
}

@article{tomekcce2024private,
  title={Private attribute inference from images with vision-language models},
  author={T{\"o}mek{\c{c}}e, Batuhan and Vero, Mark and Staab, Robin and Vechev, Martin},
  journal={Advances in Neural Information Processing Systems},
  volume={37},
  pages={103619--103651},
  year={2024}
}

@article{app10165608,
AUTHOR = {Yaghoubi, Ehsan and Khezeli, Farhad and Borza, Diana and Kumar, SV Aruna and Neves, João and Proença, Hugo},
TITLE = {Human Attribute Recognition— A Comprehensive Survey},
JOURNAL = {Applied Sciences},
VOLUME = {10},
YEAR = {2020},
NUMBER = {16},
ARTICLE-NUMBER = {5608},
URL = {https://www.mdpi.com/2076-3417/10/16/5608},
ISSN = {2076-3417},
DOI = {10.3390/app10165608}
}

@inproceedings{tricomi2024spotify,
author = {Tricomi, Pier Paolo and Pajola, Luca and Pasa, Luca and Conti, Mauro},
title = {"All of Me": Mining Users' Attributes from their Public Spotify Playlists},
year = {2024},
isbn = {9798400701726},
publisher = {Association for Computing Machinery},
address = {New York, NY, USA},
url = {https://doi-org.wwwproxy1.library.unsw.edu.au/10.1145/3589335.3651459},
doi = {10.1145/3589335.3651459},
booktitle = {Companion Proceedings of the ACM Web Conference 2024},
pages = {963–966},
numpages = {4},
location = {Singapore, Singapore},
series = {WWW '24}
}

@article{WANG2022108220,
title = {Pedestrian attribute recognition: A survey},
journal = {Pattern Recognition},
volume = {121},
pages = {108220},
year = {2022},
issn = {0031-3203},
doi = {https://doi.org/10.1016/j.patcog.2021.108220},
url = {https://www.sciencedirect.com/science/article/pii/S0031320321004015},
author = {Xiao Wang and Shaofei Zheng and Rui Yang and Aihua Zheng and Zhe Chen and Jin Tang and Bin Luo},
}

@inproceedings{feng2025agentmove,
  title={Agentmove: A large language model based agentic framework for zero-shot next location prediction},
  author={Feng, Jie and Du, Yuwei and Zhao, Jie and Li, Yong},
  booktitle={Proceedings of the 2025 Conference of the Nations of the Americas Chapter of the Association for Computational Linguistics: Human Language Technologies (Volume 1: Long Papers)},
  pages={1322--1338},
  year={2025}
}

@misc{meta2022removing,
  author       = {{Meta}},
  title        = {Removing Certain Ad Targeting Options and Expanding Our Ad Controls},
  year         = {2022},
  howpublished = {\url{https://www.facebook.com/business/news/removing-certain-ad-targeting-options-and-expanding-our-ad-controls}},
  note         = {Accessed: Jan. 2026}
}

\appendix

\section{Statistics}
\label{appendix:stat}

\begin{table}[h]
\centering
\footnotesize
\begin{tabular}{lrr}
\hline
\textbf{Demographic} & \textbf{Count} & \textbf{\%} \\
\hline
\multicolumn{3}{l}{\textbf{Gender}} \\
Male & 554 & 62.18 \\
Female & 318 & 35.69 \\
Prefer not to say & 13 & 1.46 \\
Other & 6 & 0.67 \\
\hline
\multicolumn{3}{l}{\textbf{Age Range}} \\
18--24 & 113 & 12.68 \\
25--34 & 184 & 20.65 \\
35--44 & 112 & 12.57 \\
45--54 & 112 & 12.57 \\
55--64 & 152 & 17.06 \\
65--74 & 158 & 17.73 \\
75 and over & 52 & 5.84 \\
Prefer not to say & 8 & 0.90 \\
\hline
\multicolumn{3}{l}{\textbf{Income Bracket}} \\
\$1--\$15{,}599 & 59 & 6.62 \\
\$15{,}600--\$20{,}799 & 47 & 5.27 \\
\$20{,}800--\$25{,}999 & 58 & 6.51 \\
\$26{,}000--\$33{,}799 & 55 & 6.17 \\
\$33{,}800--\$41{,}599 & 59 & 6.62 \\
\$41{,}600--\$51{,}999 & 55 & 6.17 \\
\$52{,}000--\$64{,}999 & 67 & 7.52 \\
\$65{,}000--\$77{,}999 & 56 & 6.29 \\
\$78{,}000--\$90{,}999 & 72 & 8.08 \\
\$91{,}000--\$103{,}999 & 49 & 5.50 \\
\$104{,}000--\$155{,}999 & 129 & 14.48 \\
\$156{,}000 or more & 77 & 8.64 \\
Prefer not to say & 108 & 12.12 \\
\hline
\multicolumn{3}{l}{\textbf{Education Level}} \\
Postgraduate degree level & 315 & 35.35 \\
Bachelor degree level & 366 & 41.08 \\
Year 12 or equivalent & 145 & 16.27 \\
Less than year 12 or equivalent & 45 & 5.05 \\
Prefer not to say & 20 & 2.24 \\
\hline
\multicolumn{3}{l}{\textbf{Employment Status}} \\
Employed full time & 359 & 40.29 \\
Employed part time & 199 & 22.33 \\
Unemployed and looking for work & 34 & 3.82 \\
Unemployed and not looking for work & 41 & 4.60 \\
Retired & 224 & 25.14 \\

Prefer not to say & 34 & 3.82 \\

\hline
\multicolumn{3}{l}{\textbf{Party Preference}} \\
Labor & 331 & 37.15 \\
Liberal (National Coalition) & 28 & 3.14 \\
Greens & 315 & 35.35 \\
None & 166 & 18.63 \\
Other & 51 & 5.72 \\
\hline
\end{tabular}
\caption{Filtered Users' demographics.}
\label{tab:dataset}
\end{table}

\begin{figure}[htbp]
    \centering
    \includegraphics[width=\linewidth, alt={Bar chart showing the distribution of advertisements across IAB Tier 1 categories in the filtered dataset. Retail dominates with more than 140,000 ads, far exceeding all other categories. The next largest groups include Health and Medical Services, Clothing and Accessories, Home and Garden, Food and Drink, and Business, with counts ranging between 40,000 and 70,000. Subsequent categories, such as Education, Financial Services, Arts and Entertainment, and Electronics, range between 20,000 and 40,000 ads. Many smaller categories follow with steadily decreasing counts, including Politics, Sports, Consumer Electronics, Insurance, Travel, and Real Estate. The lowest-frequency categories, each with very few ads, include Tobacco, Adult Products and Services, Weapons and Ammunition, and Non-Attributable. Overall, the distribution is highly skewed, with Retail accounting for the vast majority of ads and a long tail of niche categories.}]{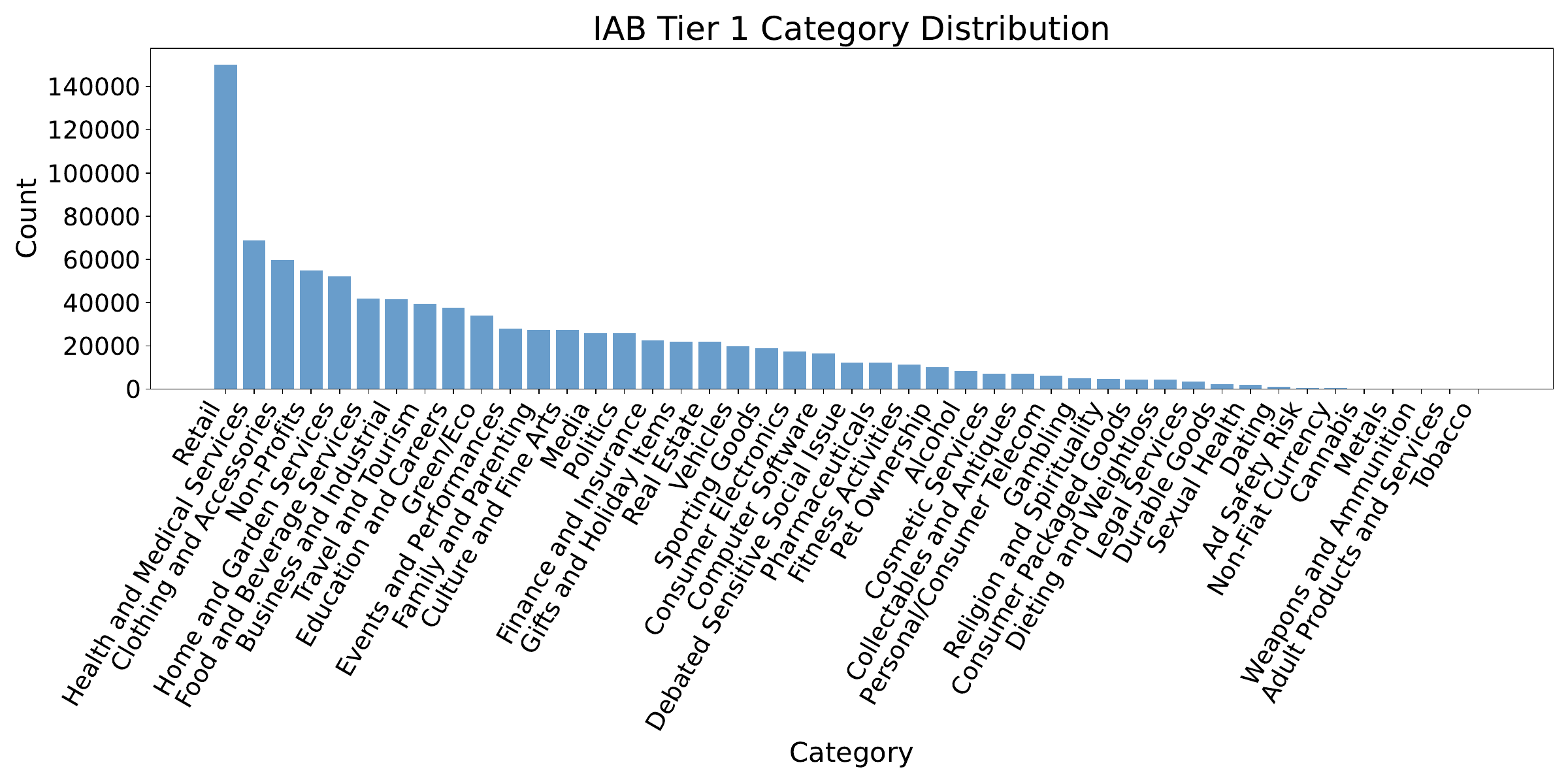}
    \caption{Category Distribution from the filtered dataset.}
    \label{fig:iab_category}
\end{figure}

\section{Parameters of Models}
\label{sec:param_of_models}
In the multimodal understanding stage, we used Gemini 2.0 Flash~\footnote{\url{https://cloud.google.com/vertex-ai/generative-ai/docs/models/gemini/2-0-flash}} with a temperature of 0.3 to allow a small degree of creativity needed for fluent yet faithful captions, while constraining outputs to a structured JSON format. For the reconstruction stages (session- and user-level), we set temperature to 0.0 to maximize determinism and reproducibility. When employing a thinking model, we used Gemini 2.5 Pro~\footnote{\url{https://cloud.google.com/vertex-ai/generative-ai/docs/models/gemini/2-5-pro}} with the default dynamic-thinking configuration enabled (thinking budget = -1). For GPT-5~\footnote{\url{https://platform.openai.com/docs/models/gpt-5}} and GPT-5 mini~\footnote{\url{https://platform.openai.com/docs/models/gpt-5-mini}}, we retained the default reasoning effort setting (``medium''). All other non-thinking models followed the same configuration as Gemini 2.0 Flash in the reconstruction stage (i.e., temperature set to 0.0).

\section{Gemini Quality Evaluation}
\label{sec:gemini_quality_eval}
\begin{table}[h!]
\centering
\footnotesize
\begin{tabular}{lc}
\toprule
\textbf{Dimension / Metric} & \textbf{Value (\%)} \\
\midrule
Caption Overall            & 99 \\
Caption Relevance            & 100 \\
Caption Correctness          & 99 \\
Caption Informativeness      & 100 \\
Descriptive Category Accuracy    & 99 \\
IAB Tier 1 Category Accuracy     & 97 \\
Key Entities Accuracy            & 99 \\
\bottomrule
\end{tabular}
\caption{Evaluation of Gemini caption quality with breakdown by sub-dimensions.}
\label{tab:caption_quality}
\end{table}

To ensure the reliability of Gemini’s structured outputs from the multimodal understanding stage, we conducted a systematic human evaluation across multiple dimensions of caption and metadata quality (Table~\ref{tab:caption_quality}). Three human annotators were involved in the process. Two annotators independently evaluated a set of 100 advertisements generated by Gemini. Each output was judged against the original advertisement image and text, assessing whether the generated caption, descriptive category, IAB Tier 1 category, and key entities accurately represented the ad content. To resolve cases of disagreement, a third annotator who had no stake or involvement in this work acted as an independent judge. For caption evaluation, given its natural language nature, we considered three sub-dimensions: \textit{Relevance}, \textit{Correctness}, and \textit{Informativeness}. A caption was marked as \textit{Overall correct} only if all three sub-dimensions were judged to be true. This adjudication procedure ensured that the evaluation results were consistent, unbiased, and suitable for supporting the accuracy of subsequent analyses and reconstruction tasks.

The evaluation results in Table~\ref{tab:caption_quality} demonstrate that our multimodal understanding stage achieved consistently high accuracy across all assessed dimensions. In particular, caption were near-perfect, with an overall correctness of 99\%. Likewise, the descriptive category, IAB Tier 1 classification, and key entity extraction all exhibited similarly strong performance. These findings indicate that the structured outputs produced by Gemini 2.0 Flash are highly robust and reliable, providing a strong foundation for the subsequent analysis and session-level and user-level reconstruction stages.

\section{Example of Human-Gemini Disagreement}
\label{sec:human-gemini_disagree}
\begin{figure}[b]
\centering
\begin{tcolorbox}[width=\linewidth, colback=gray!3, colframe=black!70, title=Human--Gemini Disagreement]

\begin{center}
  \includegraphics[width=\linewidth,height=0.15\textheight,keepaspectratio]{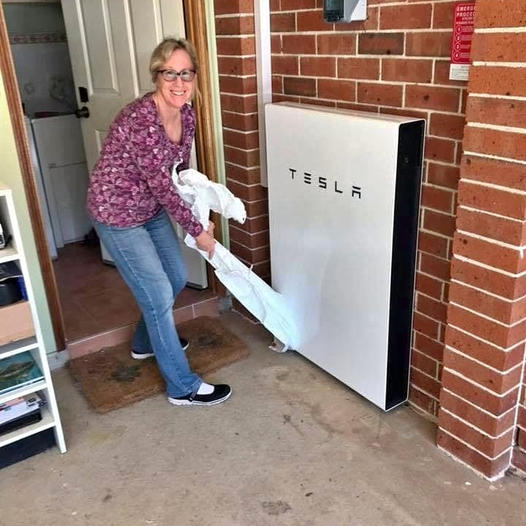}
\end{center}
\scriptsize
\textbf{Ad Title} \quad ACT Rebate Finder\\
\textbf{Ad Text} \quad We Need Homeowners Interested In Trying Solar Energy. Generate your own electricity -- you may pay nothing for it. \textbar\ Eligible participants must: \textbar\ 1. Own a home\ldots\ See more \\

\textbf{Gemini Caption}\\
The advertisement features a woman standing next to a Tesla Powerwall, a home battery system designed to store solar energy.
\hl{The woman is holding a white cloth, possibly indicating cleaning or maintenance of the unit.}
The Powerwall is installed on an exterior brick wall, suggesting a residential setting.
The slogan mentions the possibility of generating your own electricity and potentially paying nothing for it, which aligns with the purpose of the Tesla Powerwall to store solar energy for home use.
The slogan also mentions homeowners interested in trying solar energy, which is the target audience for this advertisement.

\textbf{Evaluate Caption (Yes/No for each dimension)}

\textit{Q1. Does it describe the core target/subject of this advertisement (such as the brand, product, service, event, etc.)?}\\
\checkmarkyes\ Relevant (Yes) \quad  Not Relevant (No)

\textit{Q2. Does it contain no factual errors (such as getting the categories, brands, discounts, or themes wrong)?}\\
\ Correct (Yes) \quad \checkmarkyes\ Incorrect (No)

\textit{Q3. Does it cover the key elements?}\\
\checkmarkyes\ Informative (Yes) \quad  Not Informative (No)

\textbf{Descriptive Category}  \\[0.1em] Informative, Eco-friendly, Home improvement, Clean energy promotion \\[0.1em]
\checkmarkyes\ Accurate (Yes) \quad Inaccurate (No)

\textbf{IAB Tier 1 Category} \\[0.1em] Home and Garden Services, Green/Eco, Real Estate \\[0.1em]
\checkmarkyes\ Accurate (Yes) \quad Inaccurate (No)

\textbf{Key Entities}  \\[0.1em] Tesla Powerwall, Solar energy, Home battery system, Homeowners, Electricity generation \\[0.1em]
\checkmarkyes\ Accurate (Yes) \quad Inaccurate (No)
\end{tcolorbox}
\caption{Example of a Human--Gemini disagreement on caption quality. The image and ad metadata are shown above. Human evaluation accepts the caption's topical relevance and coverage but flags a factual error (highlighted).}
\label{fig:human_gemini_disagreement}
\end{figure}

In Figure~\ref{fig:human_gemini_disagreement}, we further illustrate how human annotators evaluated Gemini's caption quality across multiple dimensions. While the model-generated caption successfully identifies the product (Tesla Powerwall), the context (residential solar energy use), and the intended audience (homeowners), the human annotator flagged a factual error. Specifically, the highlighted sentence incorrectly describes the woman in the advertisement as holding a white cloth for cleaning or maintenance, which was judged as inaccurate. This example demonstrates how human evaluation can accept the topical relevance and coverage of a caption while still rejecting specific erroneous details. Importantly, it also shows that occasional minor errors do not undermine the overall utility of the generated captions, highlighting the robustness of our evaluation framework.

\section{Temporal Segmentation and Data Cleaning Details}
\label{sec:appendix_preprocessing}

To rigorously define user sessions, given a user with a sequence of ad timestamps \( \{t_1, t_2, \dots, t_n\} \), we compute the inter-ad time intervals:
\begin{equation}
\Delta t_i = t_{i+1} - t_i, \quad \forall i \in \{1, \dots, n - 1\}
\end{equation}

Since these intervals are often highly skewed, with short gaps reflecting within-session activity and long gaps reflecting offline periods, we apply a logarithmic transformation to better capture the underlying multimodal structure:
\begin{equation}
\delta_i = \log(\Delta t_i)
\end{equation}

We then fit a kernel density estimator (KDE) over the set \( \{\delta_i\} \) to obtain a smooth estimate of the gap distribution. The KDE curve typically reveals two prominent modes:
(1) the first local maximum \( \delta_{\text{peak}} \), corresponding to frequent short gaps within sessions, and
(2) the first local minimum \( \delta_{\text{valley}} \), marking the transition to between-session gaps. We define a session threshold \( \theta \) as the midpoint between these extrema in log-space, exponentiated back to the time domain:
\begin{equation}
\theta = \exp\left( \frac{\delta_{\text{peak}} + \delta_{\text{valley}}}{2} \right)    
\end{equation}

We compute a session threshold \(\theta\) by averaging the per-user KDE-derived midpoints, resulting in a global threshold of
\(\theta = 389\) seconds. Additionally, to further prepare the ad text for subsequent processing after filtering, we remove all HTML elements and markup, resulting in clean plain-text representations free from formatting artifacts.

\section{Scalability and Economic Feasibility of the Attack}
\label{sec:scale}
To quantify the operational scalability of our threat model, we report the average computational time (52×) and cost (223×) derived from processing the full dataset with our LLM pipeline, benchmarked against the manual effort required by human annotators. We used Gemini 2.0 Flash via API for the complete pipeline (which can parallelise multiple instances of Gemini 2.0 Flash), comprising Multimodal Understanding and Session-level Reconstruction. Based on the aggregate runtime over the entire dataset, the average time required to process 100 sessions (containing approximately 680 ads on average) is 138.41 seconds, corresponding to roughly 1.38 seconds per session. Using the pricing of \$0.10 per 1M input tokens and \$0.40 per 1M output tokens for Gemini 2.0 Flash, the average inference cost for 100 sessions is \$0.179, or approximately \$0.0018 per session. In contrast, our study (Section~\ref{sec:session-level_result}) indicated that human annotators required approximately 2 hours (7,200 seconds) to analyse and annotate the same 100 sessions. At a standard wage of \$20 USD per hour, this incurs a labour cost of \$40. Comparing these metrics reveals a dramatic disparity in efficiency: the LLM approach offers a speed-up factor of approximately 52× (7,200s / 138.41s) and a cost reduction factor of 223× (\$40.00 / \$0.179). These figures show that automated adversarial profiling is both technically feasible and highly scalable, requiring minimal computational and financial resources.

\end{document}